\documentstyle[11pt,moriond,psfig]{article}

\def\AA{{\em Aston.\ Astrophys.}}
\def\APJ{{\em Astroph.\ J.}}
\def\APJL{{\em Astroph.\ J. Lett.}}
\def\MNRAS{{\em Mon.\ Not.\ R. Astr.\ Soc.}}

\def\PRD{{\em Phys. Rev.} D}

\begin{document}
\vspace*{4cm}
\title{ FRAME DRAGGING IN BLACK HOLE-PULSAR BINARIES }

\author{ N. WEX }

\address{ Max-Planck-Institut f\"ur Radioastronomie, Auf dem H\"ugel 69,\\
D-53121 Bonn, Germany }

\maketitle\abstracts{ The discovery of frame-dragging effects in binary pulsar
timing experiments requires a compact companion with sufficiently large
spin. A pulsar orbiting a fast rotating black hole could provide an
appropriate test system.  In this paper we address questions concerning the
identification of a black hole companion in such a system, the measurability
of the frame dragging caused by the rotation of the black hole, and the
measurability of the quadrupole moment, which would prove the presence of a
Kerr black hole.  }

\section{Introduction}

For radio pulsars in orbit with a compact companion, pulsar timing
observations have proved to be a powerful tool for identifying the physical
nature of the companion and studying gravitational physics present in the
binary system. The discovery of the binary pulsar PSR~B1913+16 by Hulse and
Taylor~\cite{HT75} and its continuous observation is an excellent example for
this branch of high precision astrophysics (see Joe Taylor's contribution in
this volume).  Unfortunately, perhaps the most intriguing system where such a
tool could be used, a pulsar in orbit with a black hole (BH), has yet to be
discovered. A pulsar orbiting a BH in a close orbit would certainly be a high
precision laboratory for BH physics.\cite{PT79} There are three basic
questions related to the discovery of a (tentative) BH-pulsar binary:
\begin{itemize}
\item How do we identify the BH companion?
\item What BH physics can be studied?
\item Can we prove the presence of a Kerr BH?
\end{itemize}
The answer to these three questions does not only depend on the BH properties,
but also on the properties of the pulsar and its orbit around the BH. The
observed pulsar can be a young pulsar which was formed during the second
supernova explosion in the binary system. The BH was formed during the first
explosion of the more massive star. For such a system we expect a highly
eccentric binary orbit ($\sim 0.9$) and a long orbital period (10 \dots 1000
days).\cite{LPPO94} A millisecond pulsar orbiting a BH is more difficult to
create, since we need a phase of mass-transfer to recycle the pulsar.  The
pulsar could be recycled by a low-mass companion and later be captured by a BH
in a three body interaction. The high star densities in the core of globular
clusters seem to be the most likely `breeding ground' for such kind of
systems. But also binary evolution might allow a situation where the pulsar is
created before the BH. It was argued by Ergma and van~den~Heuvel~\cite{EvdH98}
that neutron-star/BH formation is connected with other stellar parameters
besides just the mass of the progenitor star alone (at least for initial
masses $\ge$ 20 $M_\odot$). Therefore, the explosion of the more massive star
could form the pulsar, which then is recycled during the evolution of the
second star, which is still massive enough to form a BH. A millisecond pulsar
orbiting a BH is of particular interest, since the timing accuracy is
typically more than two orders of magnitude better than for young pulsars,
and, in general, millisecond pulsars are free of timing noise, at least on
time scales of a few years.

\section{Identification}

So far the best arguments for the existence of stellar mass black holes (BHs)
are based on dynamical mass estimates in X-ray binaries.  The measurement of
absorption-line velocities of the secondary star allows us to determine the
mass function of the binary, which is a lower limit to the mass of the compact
companion.  If the mass of the companion exceeds the calculated maximum mass
of a neutron star ($\sim 3 M_\odot$) we call it a BH candidate (see
the paper by Wijers~\cite{Wij96} for a list of BH candidates).

In pulsar timing experiments the mass function of the system is known with
high precision as soon as the pulsar's binary nature is identified. Even for
modest timing precision and long orbital periods, the relativistic precession
of periastron is measurable within a few years of regular timing observation
(see Fig.~1). This would provide the total mass of the system and thus,
assuming a pulsar mass of 1.4 $M_\odot$, the mass of the BH companion. However,
the absence of any mass transfer, like in X-ray binaries, does not allow (at
that moment) to distinguish between a compact object or a normal star being
the companion of the observed binary pulsar.

Additional optical or infrared observations have the potential to rule out a
massive star.  A characteristic age of the pulsar which is much larger than
the lifetime of a tentative main-sequence star companion would already
strengthen the case for a black-hole companion.  For orbital periods of less
than one year the presence of a (rotating) massive main-sequence star
companion should cause a precession of the binary orbit due to its rotation
induced quadrupole moment, like in the PSR~0045$-$7319 system.\cite{KBM+96}
This `classical' spin-orbit precession can help to rule out the presence of a
black hole, as is shown in the following section.  Depending on the size and
orientation of the orbit, the absence of eclipses or changes in dispersion
measure caused by a stellar wind can help to exclude the presence of an
extended massive star.

\begin{figure}
\centerline{\psfig{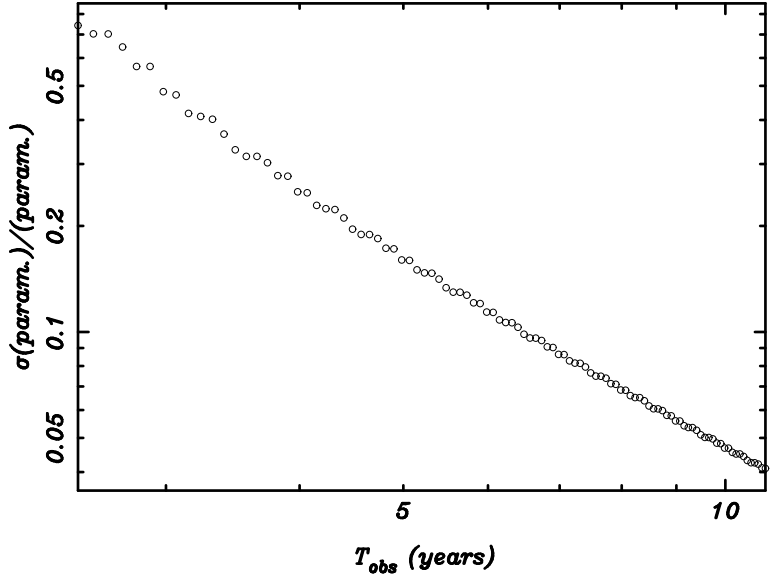}}
\caption{Simulated fractional measurement precision for the relativistic
precession of periastron, $\dot\omega$, as a function of observing time.
Estimations were done for a pulsar in a one year highly eccentric ($e=0.8$)
orbit with a 10 solar mass BH. It was assumed, that there is one timing
observation every month with a timing accuracy of 1 ms.}
\end{figure}

\section{Frame dragging}

Astrophysical BHs are expected to rotate.  The spin of the BH gives rise to a
so called gravitomagnetic field which causes the relativistic dragging of
inertial frames in the vicinity of the BH.\cite{TH85} This gravitomagnetic
field influences the motion of matter and the propagation of light in the
vicinity of the rotating body.\cite{CW95} It was suggested that in
pulsar-timing experiments a rotating BH companion may be identified by its
influence on the propagation time of the radio signals.\cite{NPS91} Based on
numerical ray-tracing calculations it was argued by Laguna and
Wolszczan~\cite{LW97} that high precision pulsar-timing experiments could,
indeed, test the gravitomagnetic field of a rotating BH companion, if the
pulsar is a millisecond pulsar in a tight orbit around a 10--20 $M_\odot$ BH
with an orbital inclination very close to 90 degrees.  It was shown, however,
by Wex and Kopeikin~\cite{WK99} that in practice a direct measurement of this
effect is not possible for stellar mass BH companions, due to the presence of
an additional relativistic effect found by Doroshenko and Kopeikin~\cite{DK95}.

A different approach in studying the importance of frame-dragging in
pulsar-timing experiments was presented by Wex and
Kopeikin~\cite{WK99}. Eighty-one years ago Lense and Thirring~\cite{LT18}
pointed out that the gravitomagnetic field of a central rotating body will
cause a precession of the orbit of a test particle (relativistic spin-orbit
coupling). In the same way, the rotation of one or both components of a binary
system will cause a precession of the binary
orbit.\cite{Bru91}$^,$\cite{BOC75} It was shown that the observation of such a
precession can lead to the direct determination of the spin of the BH
companion.\cite{WK99} In this section we give a brief summary of their basic
ideas.

Since the spin of a 10 $M_\odot$ extreme Kerr BH is more than two orders of
magnitude larger than the spin of a pulsar, the spin of the BH will completely
dominate the orbital precession, leading to a simple picture of the spin-orbit
dynamics of the binary (see Fig.~2). The precession of the angles $\phi$ and
$\psi$, which is linear in time, translates into a non-linear-in-time
evolution of the observable quantities $x$ and $\omega$, where $x$ is the
projected semi-major axis of the pulsar orbit and $\omega$ is the longitude of
periastron. In a second order approximation we can write
\begin{eqnarray}
   \Delta x_{\rm SO}(t)      &=& \dot x_{\rm SO}      \;(t-t_0)
          +\frac{1}{2}\;\ddot x_{\rm SO}      \;(t-t_0)^2 \;, \\
   \Delta \omega_{\rm SO}(t) &=& \dot{\omega}_{\rm SO}\;(t-t_0)
          +\frac{1}{2}\;\ddot{\omega}_{\rm SO}\;(t-t_0)^2 \;.
\end{eqnarray}
The observational parameters $\dot x_{\rm SO}$, $\ddot x_{\rm SO}$, $\dot
\omega_{\rm SO}$, and $\ddot \omega_{\rm SO}$ are functions of the orbital
parameters of the binary system and the spin (magnitude and orientation) of
the BH.

\begin{figure}
\centerline{\psfig{figure=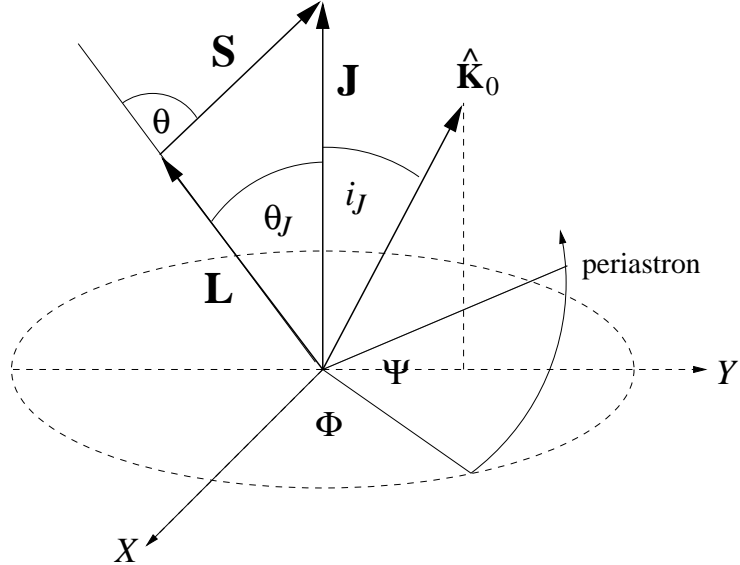,angle=-90,height=3.0in}}
\caption{
Definition of angles in the total-angular-momentum reference frame.
The invariable ($X$-$Y$) plane is perpendicular to the total angular momentum
${\bf J}={\bf L}+{\bf S}$. The line-of-sight vector ${\bf K}_0$ is in the
$Y$-$Z$ plane. The vector ${\bf J}$ is a conserved quantity and, if averaged
over one orbital period, the absolute values $|{\bf L}|$ and $|{\bf S}|$ 
are constant. Thus, the angles $i_J$, $\theta_J$, and $\theta$ are fixed. The
angles $\Phi$ and $\Psi$ change linearly with time and their precession rate is proportional to $|{\bf S}|$.
}
\end{figure}

The precession in $\omega$ caused by spin-orbit coupling is only a small
fraction of the total $\dot\omega$ ($\sim 10^{-3}$) and there is no
independent measurement of $\dot\omega_{{\rm SO}}$, until the masses of pulsar
and BH are determined with sufficient accuracy from the measurement of
additional relativistic parameters. Therefore, the first parameter indicating
relativistic spin-orbit coupling will be $\dot x$, i.e.\ a change in the
inclination of the binary orbit with respect to the line of sight (see
Fig.~3). Combined with the measurement of (total) $\dot\omega$, which allows
to derive the total mass of the system and thus gives a good estimate of the
BH mass, the measurement of $\dot x$ gives the quantity $S_\bullet \sin\theta
\sin\Phi_0$ which is a lower limit to the BH spin $S_\bullet$.  On the other
hand, if the companion is a main-sequence star and the orbit is undergoing
`classical' spin-orbit precession, then this method has the potential to rule
out a BH since general relativity predicts an upper limit for the spin of a
(Kerr) BH. Stairs {\em et al.} (this proceedings) reported the discovery of a
binary pulsar in a highly eccentric 8-month orbit, where the mass of the
companion exceeds 11 solar masses. Numerical simulations show, that if the
companion is a main-sequence star rotating at just 20\% of its breakup
velocity then even for moderate timing accuracy, the method presented here
will help to rule out a BH companion after 5 to 10 years of regular timing
observations, unless the inclination of the companion spin with respect to the
orbital plane is very small.

\begin{figure}
\centerline{\psfig{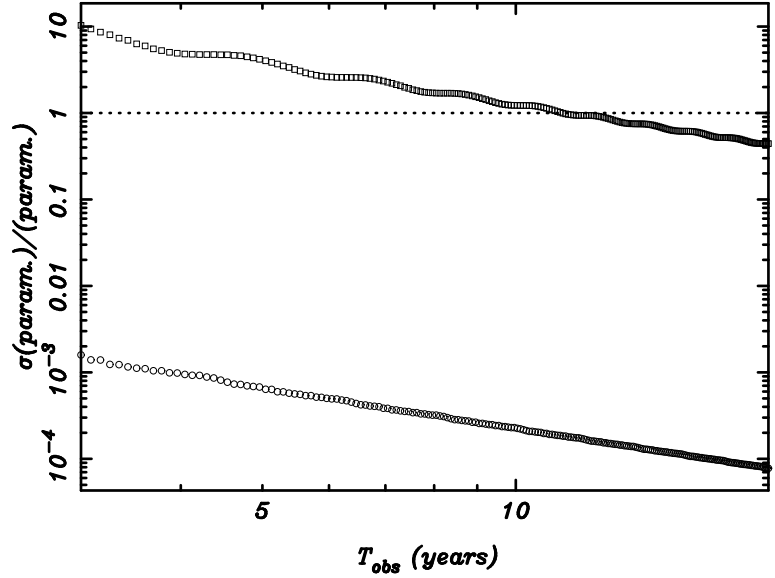}}
\caption{Simulated fractional measurement precision for the (total)
relativistic precession of periastron, $\dot\omega$ (lower points), and the
secular change, $\dot x_{SO}$ (upper points), in the projected semi-major axis
caused by frame-dragging as a function of observing time.  Estimations were
done for a pulsar in a 20 days highly eccentric ($e=0.8$) orbit with a 10
solar mass extreme Kerr BH. It was assumed, that there is one timing
observation every month with a timing accuracy of 0.1 ms (young pulsar).  This
represents a limiting case for the detection of frame-dragging in a BH pulsar
binary. A tighter orbit or/and a millisecond pulsar would certainly allow the
discovery of frame-dragging on much shorter time scales.}
\end{figure}

A millisecond pulsar BH system with orbital periods below one day
will allow the measurement of higher order precessional contributions, i.e.\
$\ddot x_{SO}$ and $\ddot\omega_{SO}$ and the separation of $\dot\omega_{SO}$,
from the total precession of periastron.  We then know the magnitude and
direction of the BH spin.

\section{Is there more?}

The `no-hair' theorem of general relativity implies that the external
gravitational field of an astrophysical (uncharged) BH is fully determined by
the mass and the spin of the black hole.  Therefore, if we are able to extract
independently the quadrupole moment of the companion from our timing
observations we could actually test whether the observed pulsar is orbiting a
Kerr BH or another compact relativistic object, like a Q or boson star. The
quadrupole moment of the BH companion will lead to an additional precession of
the binary orbit (`classical spin-orbit precession). Unfortunately these
secular changes in the orientation of the orbit caused by the quadrupole
moment of a BH companion are typically three orders of magnitude smaller than
the changes caused by the relativistic spin-orbit coupling and hence cannot be
extracted from the overall precession. On the other hand, the anisotropic
nature of the quadrupole component of the external gravitational field will
lead to characteristic short-term periodic effects every time the pulsar gets
close to the oblate companion.\cite{Wex98} Numerical simulations show,
however, that only for black hole companions that exceed $\sim1000\;M_\odot$
the quadrupole moment can give rise to an observable signature in the timing
residuals (see Fig.~4).  The discovery of such a system in our Galaxy seems,
however, unlikely.

\begin{figure}
\centerline{
\psfig{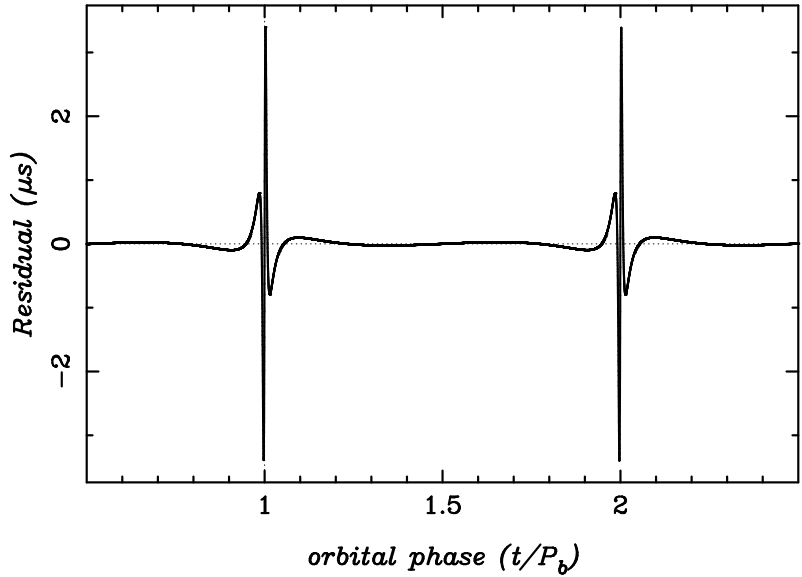}}
\caption{ Typical signature in the timing residuals caused by the
quadrupole moment of a $10^4\;M_\odot$ extreme Kerr BH companion.  We
used an orbital period of 10 days and an eccentricity of 0.9. The
inclination of the BH spin with respect to the orbital plane (the
angle $\theta$) was assumed to be 70 degrees.
}
\end{figure}

\section*{Acknowledgments}

Support from the European Union (Training and Mobility of Researchers
Programme) is acknowledged.

\end{document}